\documentclass[twocolumn,showpacs,aps,pra,superscriptaddress]{revtex4-2}
\usepackage[dvipsnames]{xcolor}
\include{def}
\usepackage{amsmath}
\usepackage{lineno}
\usepackage{xcolor}
\usepackage{xcolor}
\colorlet{RED}{red}
\usepackage{longtable}
\usepackage{multirow}
\usepackage{makecell}
\usepackage{graphicx}
\usepackage{dcolumn}

\usepackage{bm}
\include{def}

\def\be{\begin{equation}}
\def\ee{\end{equation}}

\begin{document}

\title{Does the suppression shown at LHC in O-O collisions follow
the systematics obtained for A-A collisions $?$}
\author{M.Petrovici}
\affiliation{National Institute for Physics and Nuclear Engineering - IFIN-HH, Bucharest - Romania}
\affiliation{Faculty of Physics, Doctoral School, University of Bucharest}
\author{A.Pop}
\affiliation{National Institute for Physics and Nuclear Engineering - IFIN-HH, Bucharest - Romania}
\affiliation{Hadron Physics Department}

\date{\today}

\begin{abstract}
{In this paper we present to what extent recent experimental
results 
obtained for $\pi^{0}$ suppression in O-O collisions at 
$\sqrt{s_{NN}}$=5.36 TeV fit into the systematics for much heavier 
systems. The systematics with which the comparison is made was  
published a few years ago \cite{Pet_1} in terms
of charged particles suppression $R_{AA}$ as a function of 
$\langle N_{part} \rangle$ and
$\langle dN_{ch}/d\eta \rangle$.
The values of $R_{AA}$ at 
$\langle dN_{ch}/d\eta\rangle$ and $\langle N_{part} \rangle$ 
experimentally measured and estimated by Glauber MC, respectively,
for O-O collisions
 are in good agreement with the systematics obtained for
 A-A collisions.}   
\end{abstract}
\maketitle
\maketitle

\section{Introduction}
The studies of pp collisions at LHC energies 
up to very high charged 
particle multiplicities, revealed similarities between 
pp and Pb-Pb in
the behaviour of different observables, like 
near-side long range 
pseudorapidity correlations
\cite{CMS1},  
the ($\langle\beta_T\rangle$ - $T_{kin}^{fo}$) correlation as a 
function of
charged particle multiplicity \cite{Cristi1},
azimuthal angular correlations \cite{CMS2},
geometrical scaling
\cite{Pet_2, Pet_3, Pet_4}, etc. Based on the suppression studies 
in Cu-Cu and Au-Au collisions at RHIC and Xe-Xe and Pb-Pb 
collisions at LHC and the dependence of 
$R_{AA}^N=\frac{(\frac{d^{2}N}{dp_{T}d\eta}/\langle \frac{dN_{ch}}{d\eta}\rangle)^{cen}}{(\frac{d^2N}{dp_Td\eta}/\langle\frac{dN_{ch}}{d\eta}\rangle)^{pp,INEL}}$ on $\langle dN_{ch}/d\eta\rangle$ 
\cite{Pet_1}, it was argued why even in high charged particle 
multiplicity events in pp collisions at LHC, in the limit of 
current experimental uncertainties, no suppression was 
observed, although similarities to Pb-Pb collisions for other 
observables have been highlighted.
Recently, at the LHC energy of $\sqrt{s_{NN}}$=5.36 TeV, O-O 
collsions were  measured and preliminary $R_{AA}$ results were 
reported by ALICE \cite{Stra,Mar} and CMS \cite{CMSS} 
Collaborations for $\pi^0$ and charged particles, respectively.
 In this letter we are comparing the ALICE results with the  
 results of a systematic study of  $R_{AA}$ as a function of 
$\langle dN_{ch}/d\eta\rangle$ and $\langle N_{part} \rangle$ for
A-A collisions.

\section{$\langle N_{part} \rangle$ dependence of $\langle N_{bin} \rangle$/[$\langle dN_{ch}/d\eta \rangle^{O-O}/\langle dN_{ch}/d\eta \rangle^{pp}$]}
 In the paper cited above \cite{Pet_1} was also presented a 
 dependence on 
 $\langle N_{part} \rangle$ of
 $\langle N_{bin} \rangle$/[$\langle dN_{ch}/d\eta \rangle^{A-A}/\langle dN_{ch}/d\eta \rangle^{pp}$] for Cu-Cu and Au-Au at 
 $\sqrt{s_{NN}}$=200 GeV, Xe-Xe at $\sqrt{s_{NN}}$=5.44 TeV, 
 Pb-Pb at $\sqrt{s_{NN}}$=2.76 TeV and 5.02 TeV using for
 $\langle dN_{ch}/d\eta \rangle$ the values reported in 
\cite{STAR_dNchdeta, BES_CuCu_dNchdeta, XeXe_data, 
PbPb_276_dNchdeta, PbPb_502_dNchdeta}, $<N_{bin}>$ estimated within 
the Glauber MC approach \cite{Loiz_1} and the 
$\langle \frac{dN_{ch}}{d\eta} \rangle)^{pp}$ INEL values at the 
corresponding collision energies (Fig.1). Within the error bars, a 
good scaling is observed. The rather large values of the error bars
 are due to uncertainty in $\langle N_{bin} \rangle$ estimates 
 \cite{Loiz_1}. In Fig.1  are represented by yellow squares the 
 values corresponding to O-O collisions at 
$\sqrt{s_{NN}}$=5.36 TeV for different centralities, using 
$\langle dN_{ch}/d\eta \rangle$ reported by the ALICE Collaboration
 \cite{Stra, Mar}. $\langle N_{part} \rangle$ and
$\langle N_{bin} \rangle$ estimated within the Glauber MC approach 
were taken from \cite{Loiz} for the corresponding centralities. As 
can be observed, the values for O-O collisions
follow the systematics obtained in heavy ion collisions.
 
\begin{figure} [h]
\begin{center}
\includegraphics[width=0.90\linewidth]{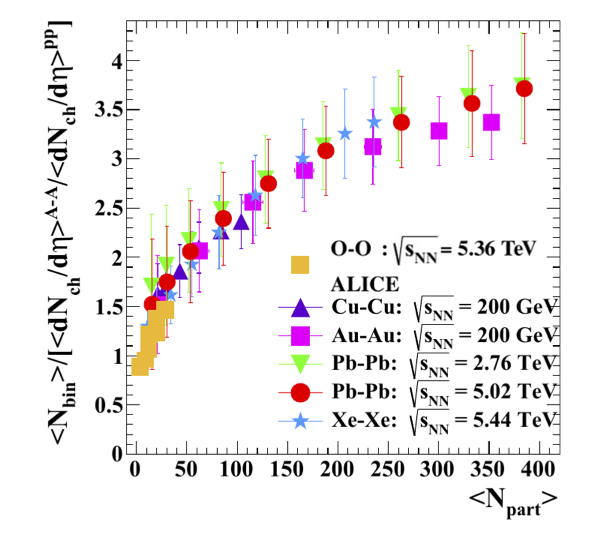}
\caption{
$\langle dN_{bin} \rangle$/[$\langle dN_{ch}/d\eta \rangle^{A-A}/\langle dN_{ch}/d\eta \rangle^{pp}$] values 
in O-O collisions at $\sqrt{s_{NN}}$=5.36 TeV are presented by yellow 
squares using recently  
measured $\langle dN_{ch}/d\eta \rangle$ for different centralities 
by the ALICE Collaboration \cite{Stra, Mar} on top of Fig.14 from 
\cite{Pet_1}}
\end{center}
\end{figure}
\begin{figure} [b]
\begin{center}
\includegraphics[width=0.90\linewidth]{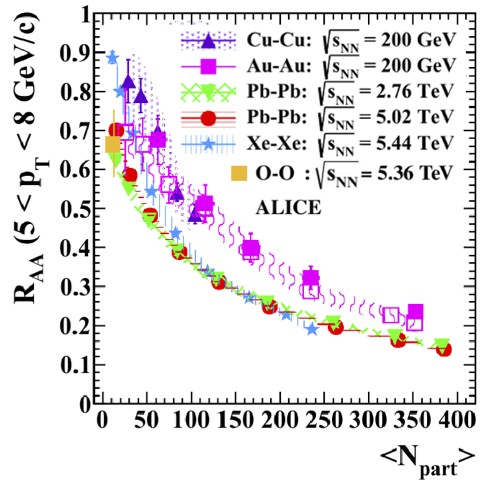}
\caption{$R_{AA}$ as a function of $\langle N_{part} \rangle$ from 
\cite{Pet_2} over which the corresponding experimental value for O-O 
is represented by the yellow box.}
\end{center}
\end{figure}

\section {$R_{AA}$ - $\langle N_{part} \rangle$  dependence} 
As it was presented in a previous paper \cite{Pet_2}, the transverse
overlap area of the nucleons suffering more that single collisions 
$S_{\perp}^{core}$ shows a very good scaling as a function of 
$N_{part}$, both being estimated within the Glauber MC approach for 
the Au-Au at energies measured at RHIC and Pb-Pb at energies measured
 at LHC. Similar results were obtained for $S_{\perp}^{var,core}$ for
  Cu-Cu, Au-Au at top energy at RHIC and Xe-Xe and Pb-Pb at energies 
  measured at LHC. The aspect ratios of the transverse overlap area 
  $S_{\perp}$ for different systems and energies are also rather 
  similar. 
With the assumption that the average path length of a parton traveling 
the deconfined zone $L^2\propto S_{\perp}$ one could expect a good scaling of the average suppression as a function of $N_{part}$ at a given collision energy. 
This was presented in ref. \cite{Pet_1}. 
For O-O at $\sqrt{s_{NN}}$=5.36 TeV, an estimate of 
$N_{part}$=10.8 for 0-100\% centrality based on the Glauber MC 
approach was reported in a recent publication \cite{Loiz}. In Fig.2 
by the yellow box is represented the $R_{AA}$ value reported as 
preliminary results by the ALICE Collaboration \cite{Stra, Mar}  at 
the corresponding $N_{part}$ on top of the compilation presented in 
\cite{Pet_1}. The agreement between the O-O experimental result and 
the $N_{part}$ dependence of $R_{AA}$ obtained in Xe-Xe and Pb-Pb at 
similar collision energies is very good.

\section{$R_{AA} - \langle dN_{ch}/d\eta \rangle$ dependence}

 Within the phenomenological parton energy loss approach, 
the fractional energy loss is given by:
 
 \begin{equation}
 \frac{\Delta E}{E}\propto T^aL^b
 \end{equation} 
\noindent
 where a and b take different values \cite{Xu}. This could explain 
 the collision energy dependence of $R_{AA}$ as a function of   
 $N_{part}$, the temperature being rather different as far as
     $\langle dN_{ch}/d\eta \rangle$ is different for a given value 
     of $N_{part}$ \cite{Pet_1}. Based on this argument, in 
     \cite{Pet_1} we also presented $R_{AA}$ as a function of 
     $dN_{ch}/d\eta$ for the same systems. In such a representation 
     the energy dependence of the scaling does not appear anymore 
     although the aspect ratio of 
     $S_{\perp}$ for different systems and collision energy is rather 
     different. The study of the azimuthal dependence of $R_{AA}$ done
      by the PHENIX Collaboration for Au-Au collision at $
      \sqrt{s_{NN}}$=200 GeV and 30-40\% centrality showed a larger
       suppression out of the reaction plane relative to the one 
       in-plane \cite{PHE}. It seems that this effect is less important 
       in the average value of $R_{AA}$ and the temperature dependence 
       of the energy loss in the deconfined matter plays the main 
       role. 
     
     \begin{figure} [h]
\begin{center}
\includegraphics[width=0.90\linewidth]{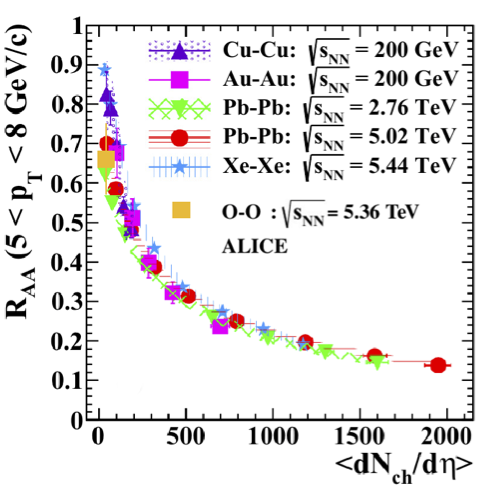}
\caption{$R_{AA}$ as a function of $\langle dN_{ch}/d\eta \rangle$ 
from \cite{Pet_2} on which the corresponding experimental value for 
O-O is represented by the yellow box.}
\end{center}
\end{figure}
     
     Taking the measured value of $\langle dN_{ch}/d\eta \rangle$ for 
     O-O collisions at 
$\sqrt{s_{NN}}$=5.36 TeV reported by the ALICE Collaboration 
\cite{Stra, Mar} , the result is presented in Fig.3 by the yellow 
square on top of the compilation presented in Fig.8 of reference 
\cite{Pet_2}. 
The $R_{AA}$ for the O-O case is in good agreement with the results 
obtained for much heavier systems at RHIC and LHC at the same 
$\langle dN_{ch}/d\eta \rangle$.

\section{Conclusions}
Although the alpha-like cluster structure of the colliding 
O nuclei could 
result in peculiar behaviour of many observables, it seems that the 
average value of $R_{AA}$ is in very good agreement with those 
extracted 
from Pb-Pb and Xe-Xe collisions at a similar collision energy at the 
same 
$\langle dN_{ch}/d\eta \rangle$ or 
$\langle N_{part} \rangle$ values.
The expected results from detailed multi differetial data analysis of 
the information 
collected in alpha-like light ion collisions at LHC will show to which
 extent 
the previously established systematics in heavy ion collisions and pp
 at LHC energies
\cite{Pet_1, Pet_3, Pet_4, Top} are also followed  in the case of such
 collisions.

\bibliography{Bib}

\end{document}